# Mn$_4$N ferrimagnetic thin films for sustainable spintronics

T. Gushi[1,2], M. Jovičević Klug[3,‡], J. Peña Garcia[3], H. Okuno[4], J. Vogel[3], J.P. Attané[2], T. Suemasu[1], S. Pizzini*[,3], L. Vila[†,2]

1. Institute of Applied Physics, Graduate School of Pure and Applied Sciences, University of Tsukuba, Tsukuba, Ibaraki 305-8573, Japan

2. Univ. Grenoble Alpes, CEA, CNRS, Grenoble INP, INAC, SPINTEC, F-38000 Grenoble, France

3. Univ. Grenoble Alpes, CNRS, Institut Néel, F-38042 Grenoble, France

4. Univ. Grenoble Alpes, CEA, INAC, MEM, F-38000 Grenoble, France

*stefania.pizzini@neel.cnrs.fr

† Laurent.vila@cea.fr

‡Permanent address : Kiel University, Institute for Materials Science, Kaiserstraße 2, 24143 Kiel, Germany

**Spintronics, which is the basis of a low-power, beyond-CMOS technology for computational and memory devices, remains up to now entirely based on critical materials such as Co, heavy metals and rare-earths. Here, we show that Mn$_4$N, a rare-earth free ferrimagnet made of abundant elements, is an exciting candidate for the development of sustainable spintronics devices. Mn$_4$N thin films grown epitaxially on SrTiO$_3$ substrates possess remarkable properties, such as a perpendicular magnetisation, a very high extraordinary Hall angle (2%) and smooth domain walls, at the millimeter scale. Moreover, domain walls can be moved at record speeds by spin polarised currents, in absence of spin-orbit torques. This can be explained by the large efficiency of the adiabatic spin transfer torque, due to the conjunction of a reduced magnetisation and a large spin polarisation. Finally, we show that the application of gate voltages through the SrTiO$_3$ substrates allows modulating the Mn$_4$N coercive field with a large efficiency.**

**TEXT**

The development of artificial intelligence and big data requires the development of high-speed and low-power memories and processors. In this context, spintronics, which aims at using the electron spin to carry and manipulate data, possesses compelling advantages over competing technologies: intrinsic non-volatility, room-temperature operation, and compatibility with the CMOS technology. It has thus been shown to possess the potential to disrupt not only the memory market, with the on-going commercialization of non-volatile magnetic memories such as MRAMs, but actually the whole information technology market, through the creation of post-CMOS [1] or of neuromorphic technologies[2].

Spintronics Achilles' heel might be its hazardous dependence on cobalt, rare-earth elements, and heavy metals (especially tungsten and platinum). These materials are indeed based on elements identified as critical by the government agencies of developed countries, because of the likelihood and impact of supply shortfalls, and of various geopolitical and environmental factors[3,4].

In this work, we will describe the magnetic properties of a new material, made of abundant and cheap elements, and show that it is a promising candidate for the development of a sustainable spintronics: epitaxial $Mn_4N$ films. Beyond features such as a perpendicular magnetisation, or a large Extraordinary Hall Effect (EHE), we show that in this material record Domain Walls velocities can be obtained under current at room temperature. Whereas in the past years the whole spintronics community shifted its focus from spin-transfer torques to spin-orbit torques, these results shows that classical spin-transfer torques remains highly competitive for current-induced DW motion. Also, we show that the coercivity of $Mn_4N$ films can be easily tuned by a gate voltage, this ability constituting a tool for DWs, nanomagnets or skyrmions manipulation.

**Growth and crystal structure**

$Mn_4N$ is one of the few known magnetic nitrides. While bulk magnetic nitrides have been studied in the 1960's and 1970's[5], only a few groups have succeeded in growing magnetic nitrides thin films, on MgO or $SrTiO_3$ (STO) substrates, using molecular beam epitaxy[6,7], pulsed laser deposition[8] or magnetron sputtering[9]. They grow in the anti-perovskite structure depicted in Figure 1a, with a central N atom, surrounded by metal atoms located on two inequivalent atomic sites, and anti-ferromagnetically coupled.

These films present a very large technological interest: they are made of only abundant elements, which responds to the increasing need for sustainable and low cost electronics. They are ferrimagnets with a low spontaneous magnetisation Ms, which should allow increasing the efficiency of spin transfer and spin orbit torques. This should facilitate magnetisation manipulation, increase the switching speed and lower the power consumption of spintronic devices. The small spontaneous magnetisation also provides a relatively small sensitivity to externally applied magnetic fields: ferromagnetic-based devices do not need to be magnetically shielded when embedded in memory or computational devices, and the production of stray fields is minimized.

Among the magnetic nitride series, $Mn_4N$ holds a characteristic that is much looked for in spintronics: in appropriate growth conditions, it presents a perpendicular magnetisation. This property, associated to the low $M_s$, allows foreseeing high current-driven domain wall speeds by spin transfer torques[10].

Here, 10 nm thick films were grown epitaxially on an STO substrate. As recently observed[11], the small lattice mismatch between the film and the substrate allows obtaining a high crystalline quality. The X-Ray Diffraction (XRD) spectra of Figure 1b demonstrates the epitaxial growth on the STO substrate with a (001) texture. The quality of this epitaxial growth is also evidenced by $\omega$-scan rocking curves and RHEED diffraction (cf. Methods). The typical RMS roughness of the layer, measured by Atomic Force Microscopy, is smaller than 0.7 nm. Scanning Transmission Electron Microscopy images of the layer are shown in Figure 1c and 1d, they further evidence the very high crystalline quality, with a perfect epitaxy, free of any dislocation of the Mn$_4$N epilayer and its sharp interface with the Ti terminated STO.

**Magnetic properties**

The magnetic properties of the films were studied by Vibrating Sample Magnetometry, SQUID and transport measurements. The extracted spontaneous magnetisation is small (M$_S$=6.6 x 10$^4$ A/m), whereas the uniaxial magnetic anisotropy is large enough (K$_u$= 1.1 x 10$^5$ J/m$^3$) to obtain perpendicular magnetisation. The hysteresis loop measured by extraordinary Hall effect (Figure 1e) is perfectly square, with 100% remanence. As the magnetic field acts on the magnetisation through a $-\mu_O \vec{H}.\vec{M}$ Zeeman energy density, one can expect this material, which possesses a small M$_s$, to possess a large coercive field. The coercivity remains however quite low (170 mT): as the crystalline quality is high, there are few DW pinning sites.

Also, magneto-Optical Kerr microscopy observations of the magnetic domain structure have been made for partially reversed states (cf. Figure 1.f). The low pinning results in remarkably huge domains, at the millimeter scale, with smooth and very long DWs[11].

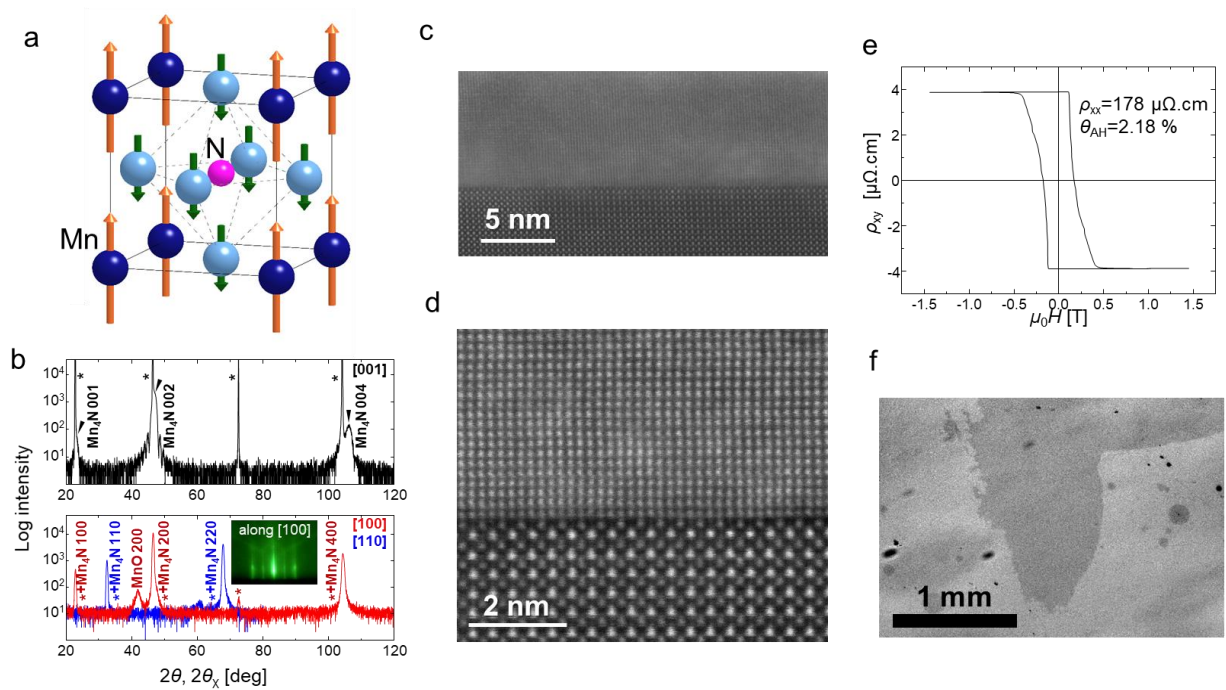

**Figure 1 | Structural and magnetic properties of Mn$_4$N ferromagnetic thin films** (a) Scheme of the anti-perovskite structure of Mn$_4$N films. The metallic ions located on the corners and at the centers of the faces are antiferromagnetically coupled. (b) Out-of-plane (top) and in-plane (bottom) XRD spectra of a Mn$_4$N(10 nm)/STO deposit, with Kiessig fringes testifying of the crystalline quality. The in-plane XRD measurements were performed by setting the scattering vector along STO[100] (red) and STO[110] (blue). Asterisks correspond to the STO substrate diffraction peaks. (c) Scanning Transmission Electronic image of the Mn$_4$N/STO sample corroborating the crystalline quality of Mn$_4$N. (d) Higher magnification image showing the perfect epitaxy of the Mn$_4$N with the Ti terminated STO interface and its sharpness, any dislocation is observed on the TEM specimen. (e) Hysteresis loop measured by Extraordinary Hall effect, for an applied field perpendicular to the surface. (f) Magneto-optical Kerr microscopy observation of the sample in a partially reversed state of magnetisation (M/M$_S$=-0.1). The reversed domain appears in black.

## Quasi-static transport measurements

The resistivity of the layers is around 180 μΩ cm at room temperature, comparable to CoFeB-based materials. Although $Mn_4N$ is made of only light elements, its Extraordinary Hall Effect angle is large (around 2%), comparable to that of materials with spin-orbit coupling such as FePt or TbCoFe[12].

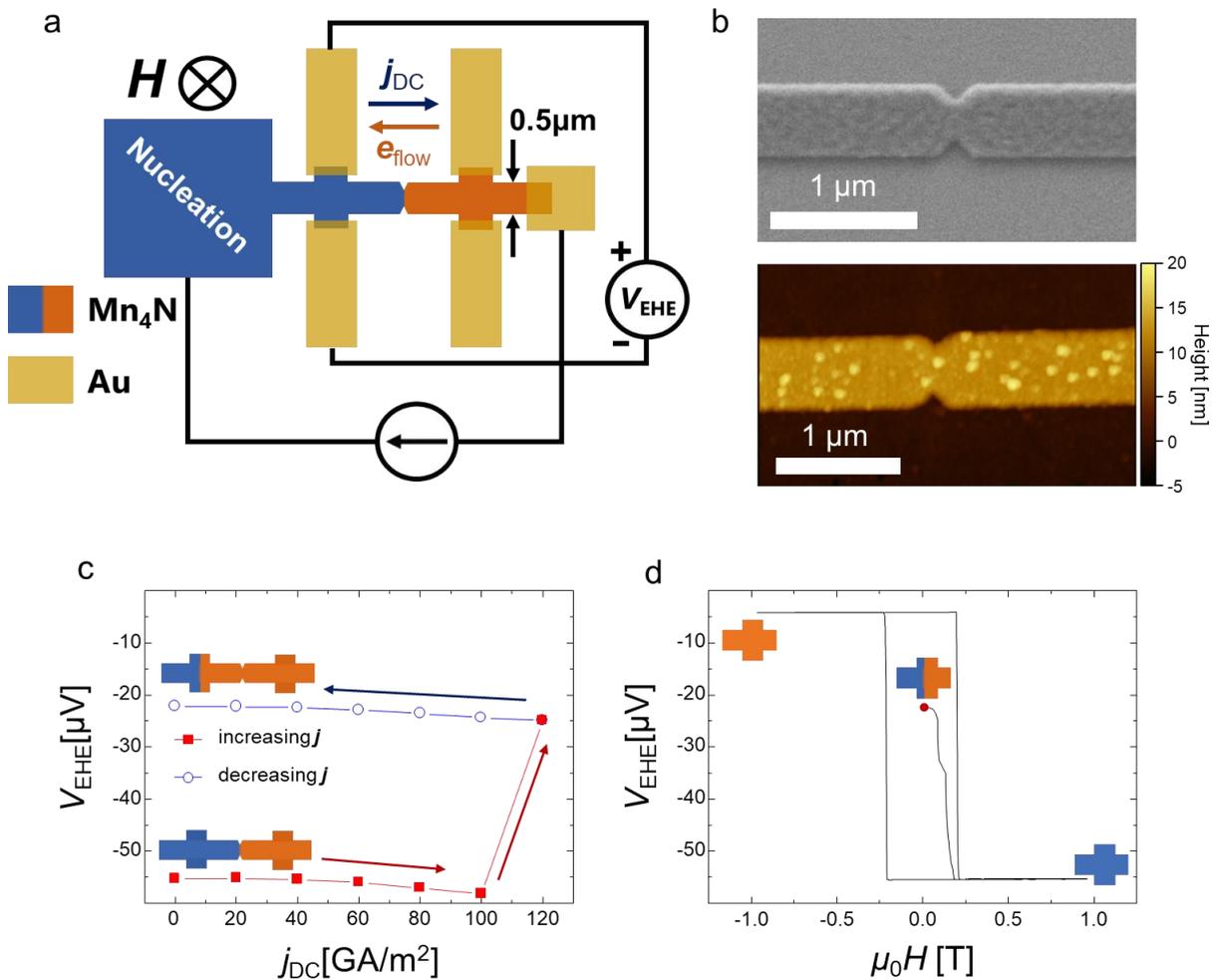

**Figure 2| Magneto-transport measurements** (a) Schematics of the device and of the probe connection (b) SEM images (top) of the constriction before etching, and AFM image (bottom) after etching. (c) Room temperature Hall voltage recorded at the left Hall cross as a function of the applied current density. The DW is firstly pinned on the constriction. It is then depinned by the current, and finally propagates within the left Hall cross, causing a partial reversal (64%) of the magnetisation. (d) Hall voltage of the first Hall cross obtained when increasing the field from 0 to positive values, the starting point being the final state of fig. (c). This confirms that after the application of the DC current in (c), the DW is indeed located in the Hall cross.

The efficiency of the spin-transfer torque was first studied by extraordinary Hall measurements, using DC currents and quasi-static fields in the device geometry depicted in Figure 2a. As seen in

Figure 2b the Mn$_4$N layer was patterned by ion milling into 0.5 µm wide Hall bars, with a constriction patterned in between the two Hall crosses (cf. § Methods). The patterning induces a sharpening of the switching, with very square hysteresis loops (cf. Figure 2d). The coercivity increase to 0.2 T is mostly due to the difficulty to nucleate, the minor loops showing that the threshold field for DW propagation remains low, around 0.15 T. The typical coercive field can be compared to the critical current density required to induce DW motion. As seen in Figure 2c, the application of a DC current allows inducing the DW motion at zero field, for a current density of 1.2x10$^{11}$ A/m$^2$. The DW is depinned from a notch, and propagates over more than 5 µm before getting pinned at the cross. The spin transfer efficiency, *i.e.*, the ratio between the typical coercive field and the DC current required to induce DW motion, is thus around 1.25×10$^{-12}$ Tm²/A. This value is two orders of magnitude higher than that of NiFe, and similar to that of systems with heavy elements and large spin-orbit coupling such as FePt[13], or of TbCoFe and CoGd ferrimagnets[14].

Note that in most magnetic thin films studied in the last years, the DWs were moved by the spin-orbit torque arising from the growth on a Spin-Hall effect material, and from the presence of the interfacial Dzyaloshinskii-Moriya interaction (DMI) which favours chiral Néel walls[15,16,17]. On the contrary, DW motion is driven here by the classical Spin Transfer Torque (STT): since the Mn$_4$N layer is thick (10 nm), interfacial effects should be negligible, even in the presence of an interfacial DMI. The DW propagates in the direction of the electron flow, which is consistent with this interpretation. In order to rule out the contribution of spin-orbit torques to the DW dynamics, we carried out Kerr microscopy measurements of domain wall motion under in-plane magnetic field, which demonstrated the absence of DMI in our system (cf. Supplementary information).

**DW velocities**

The efficiency of the spin transfer torque was also investigated by measuring directly current-driven DW velocities by Kerr microscopy. A 10 nm thick Mn$_4$N layer was patterned into 10 µm (or 20 µm) long, 1 µm (resp. 2 µm) wide strips. Twenty wires were connected in parallel via a large injection pad, in order to optimize the impedance matching with the voltage generator, and to obtain a statistically relevant measurement of the DW velocity (cf. Figure 3a).

DWs can be created using an out-of-plane field pulse, leading to the magnetic configuration of Figure 3a, with one DW per wire. Current pulses of durations ranging from 1 to 3 ns, with current densities up to 1.5x10$^{12}$ A/m$^2$ were then applied to induce DW motion. The differential image of Figure 3b allows measuring the distance over which the DWs have propagated during the pulse, therefore giving access to the average DW velocity. The limited dispersion of the DW displacements among the nanowires points out that for high current densities DWs are weakly pinned by defects (*i.e.*, the behaviour is not stochastic).

Figure 3c shows the variation of the domain wall velocity as a function of the current density. The behaviours are nearly identical for 1µm and 2µm wide strips. For low current densities J, in the thermally activated regime where the DW movement is hindered by defects, the speed changes exponentially with J. Above around J=8.5x10$^{11}$ A/m$^2$, the DW speed starts varying linearly with J, and very high DW velocities (up to 900 m/s) are reached for 1.3x10$^{12}$ A/m$^2$, in the direction of the electron flow.

Such high DW velocities deserve some discussion. Current driven domain wall motion by STT has been observed in thin films with both in-plane[18,10] and out-of-plane magnetisation. In in-plane magnetized films of permalloy, the measured DW velocities are typically of the order of 100 m/s, for currents densities around $1.5 \times 10^{12}$ A/m². However, in these systems the DW motion is largely affected by pinning, and in general a wide dispersion of DW displacements is observed even in the flow regime[19].

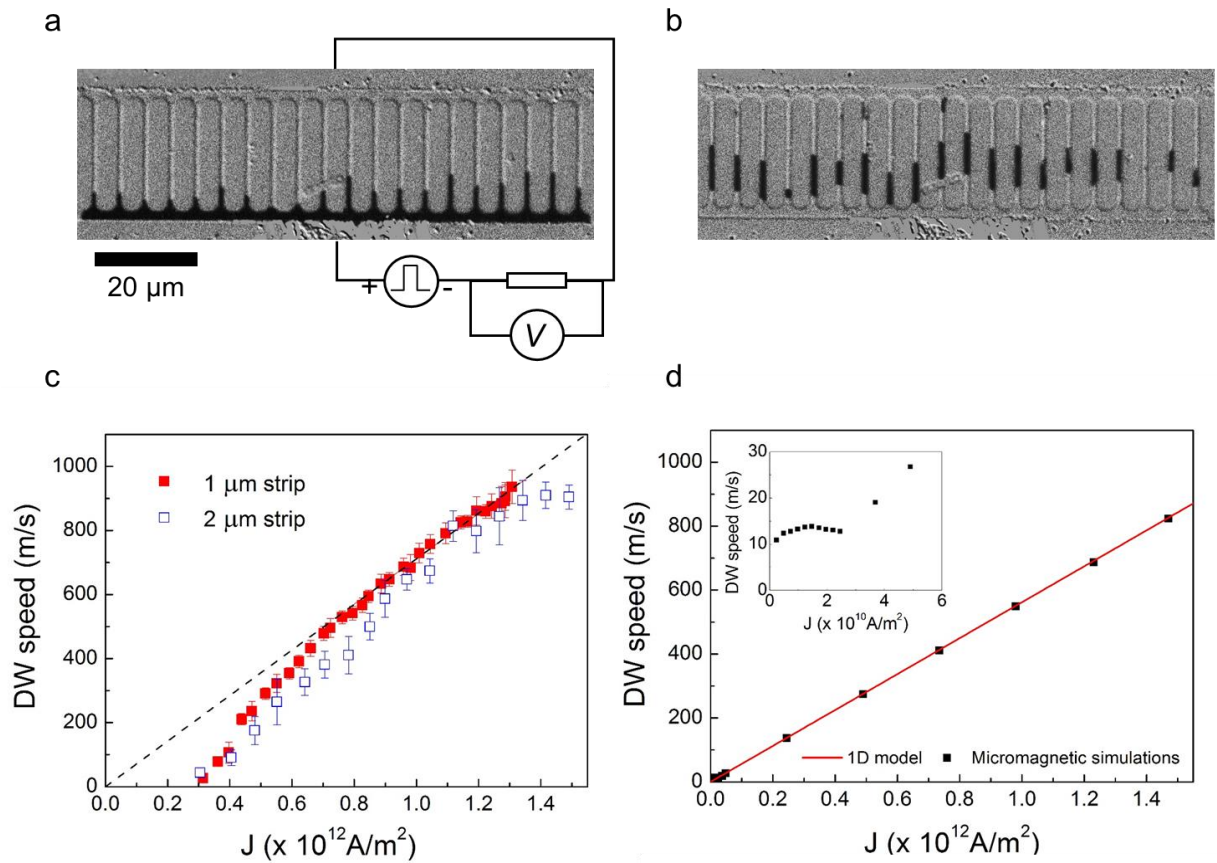

**Figure 3 | Current induced domain wall motion experiments** (a) MOKE microscopy image of the wires, and scheme of the electrical setup. A partially reversed state is achieved by applying a short field pulse. The reversed domain appears in black. (b) Differential image after the application of 8 pulses of 1 ns duration, with a current density of $+1.55 \times 10^{12}$ A/m². The black area corresponds to the zone over which the DW moved during the pulse. Note that some nucleations of reverse domains appear for such large current densities. (c) Domain wall velocity vs. current density measured for 1 and 2 µm-wide strips. The domain wall velocity reaches an average above 900 m/s at $1.3 \times 10^{12}$ A/m². The dotted line is a guide to the eye to emphasize the linear regime. (d) Theoretical DW velocity (red line) obtained using the 1D model (equation 3) using the experimental magnetic parameters ($M_s$ and $K_u$) and taking P=0.7 and α=0.03. The black dots are the results of micromagnetic simulations using the same parameters. The inset is a zoom of the low current density regime.

Moreover, since DWs are wide (>100nm) in these in-plane systems, the interest for applications to logic and storage devices is limited. Quite high spin transfer efficiencies have been measured in systems with perpendicular magnetic anisotropy (PMA) such as Co/Ni or FePt[20], but in these systems DW motion at zero field is usually hindered by pinning, which induces a large coercivity and a stochastic behaviour. In Pt/Co/Pt, DWs could not be moved by STT, due to the limited spin polarisation within the Co layer[21]. On the other hand, very high DW mobilities were found in magnetic semiconductor films such as (Ga,Mn)(As ,P) for very low current density[22,23]. However these measurements were done at low temperature, and the current density was limited to some $10^{10}$ A/m$^2$, with a maximum speed of 30m/s at $10^{10}$ A/m$^2$.

The velocities observed in Mn$_4$N can thus be considered as a milestone in the history of spin transfer torques. Although in recent years the community has focused nearly entirely on SOT-driven DW dynamics, the giant DW velocities obtained in Mn$_4$N, show that the STT can be competitive with SOT. SOT has been found to be an efficient mechanism to drive DWs in non-centrosymmetric multilayers, in which a FM is deposited on a heavy metal like Pt[16,17]. The prototypical example is Pt/Co(0.6nm)/AlOx[24] where the chiral Néel structure acquired by the DWs in the presence of interfacial Dzyaloshinskii-Moriya interaction[15] leads to a high mobility (v=400m/s for J=3×10$^{12}$ A/m$^2$) with a satisfying reproducibility of the DW movements. Higher SOT-driven DW mobilities were obtained recently in a multilayer structure in which two Co/Ni layers were coupled anti-ferromagnetically through a Ru layer[25], or in GdCo ferrimagnet layers deposited on top of Pt[26]. In these systems the DW velocity was clearly observed to vary with the total magnetisation. In the former experiment, a maximum velocity of v=750 m/s for J=3x10$^{12}$ A/m$^2$ was obtained at room temperature using a Ru spacer layer, so that the stack magnetisation was reduced to 8% of the spontaneous magnetisation of a single Co/Ni layer. In the latter one, a peak velocity of 1250 m/s was reported at 240 K (the angular moment compensation temperature) for a current density of 2x10$^{12}$ A/m$^2$.

In this context, the velocities presented here at room temperature for DWs in Mn$_4$N are comparable to the best results obtained using SOTs. Moreover, they represent the first demonstration of very efficient current-driven DW motion by pure STT using a system with PMA.

**Origin of the high domain wall mobility**

In order to understand the microscopic origins of the high current-induced DW velocities achieved in Mn$_4$N thin films, we have considered the analytical expression of the STT in the adiabatic limit. In a perfect nanowire with out-of-plane magnetisation, the 1D model describes well the main features of the DW velocity under a spin-polarised current[27,28].

Magnetisation dynamics is governed by the Landau-Lifshitz-Gilbert equation, with additional terms taking into account the effect of spin-polarised current on the magnetisation. In our system with PMA and negligible Dzyaloshinskii-Moriya interaction, the DWs are expected to be in the Bloch configuration at rest (see Supplementary Information). In the adiabatic limit of current-driven dynamics, the DW can move continuously only when its magnetisation can start precessing in order to align with the spin polarisation of the incoming conduction electrons (*i.e.*, along z in a system with PMA). To do so, the DW energy has to overcome the anisotropy energy $K_D$ given by the difference in energy between the Bloch and the Néel DW configurations. This occurs above a threshold critical current density $J_c$:

$$J_c = \frac{2e}{\hbar P}\Delta(\varphi)K_D, \qquad (1)$$

where P is the spin polarisation of the conduction electrons, $\Delta(\varphi)$ is the DW width as a function of the tilt angle $\varphi$, the angle between the magnetisation in the center of the DW and the DW plane. The anisotropy energy $K_D$ depends on the sample geometry and on the spontaneous magnetisation (see Supplementary Information). Using the experimental strip geometry (1 µm wide and 10nm thick) and the material parameters measured experimentally for the Mn$_4$N film ($M_s = 6.6 \times 10^4$ A/m, $K_u = 0.11 \times 10^6$ J/m³), and taking $A = 10$ pJ/m, P = 0.7, the critical current density is expected to be $J_c \approx 1.9 \times 10^{10}$ A/m². This is much lower than that of permalloy strips with in-plane magnetisation ($J_c \approx 10^{13}$ A/m²)[10].

While a non-adiabatic torque has been invoked to explain the DW motion observed experimentally in permalloy strips below $J_c$, in our case the observed DW motion may be attributed to the adiabatic torque alone. [29,30]

As seen in the Supplementary Information, when $J \gg J_c$ the adiabatic torque drives the DWs with a velocity given by:

$$\boldsymbol{v} \approx \frac{1}{1+\alpha^2}|\boldsymbol{u}| \qquad (2)$$

where $\boldsymbol{u} = \frac{g\mu_B}{2eM_s}P\boldsymbol{J}$ is the spin-drift velocity, parallel to the electron flow.

From Equation 2, the mobility $v/J$ is proportional to the ratio P/M$_s$ of the spin polarisation and the spontaneous magnetisation, with a negligible dependence on the damping parameter $\alpha$ that is expected to be much smaller than unity. The spin polarisation P is then the only fitting parameter and an estimation of its value in Mn$_4$N can be obtained from the comparison between the experimental DW velocities and the result of Equation 2 in which P is varied. Experimentally, a linear regime indeed appears in the speed versus current density curve above the thermally activated regime. The best agreement between experiments (Figure 3c) and 1D model (Figure 3d) is obtained using a polarisation value around P=0.7, close to the polarisation of the density of states obtained using first-principle calculations[31]. We can then conclude that the large domain wall speeds observed for Mn$_4$N are due to the conjunction of the small value of the spontaneous magnetisation and of the high spin polarisation.

In order to confirm the different features predicted by the 1D model we have also performed micromagnetic simulations (see Supplementary Information). The results reported in Figure 3d show an overall agreement of the 1D model with the 2D simulations, except for minor differences at low **J** values.

Note that in our simulations we have neglected the effect of the non-adiabatic torque on the DW velocity. While this term had to be introduced to explain the DW dynamics in in-plane magnetized systems[18], it has actually a much lower influence on the DW velocity our system.

Below the critical current density $J_c$, the presence of the non-adiabatic torque results in a steady regime motion, with a velocity $v = \frac{\beta}{\alpha}u$. This effect is not observable in our system, as the critical current $J_c$ is extremely small and obscured by the creep regime. Above $J_c$, in the precessional regime, a term $\frac{\alpha\beta}{1+\alpha^2}u$ has to be added to the velocity of Equation 2. However, since $\alpha$ is expected to be much smaller than unity, this extra term is negligible with respect to the adiabatic term. Although the measurement of the non-adiabatic term has been the subject of severe debates[32], its main effect is to explain the presence of DW motion below the critical current. Here we show that the adiabatic torque alone can explain the measured DW speeds down to low current densities, because of the small critical current densities characteristic of PMA systems with small $M_S$. This point is further discussed in the Supplementary Information, where we also study the hypothesis according to which $\alpha = \beta$.

**Electric field effects**

The last property we would like to address is the effect of the electric field on the anisotropy of the Mn$_4$N layer. A back gate voltage was applied to the bottom of the 300 µm STO substrate, while measuring the coercivity of a 4.2 nm-thick Mn$_4$N layer with a magnetic field (cf. § Methods). The measurements were performed at 10 K, so that the dielectric constant of STO reaches very high values (30 000). The gate voltage acts on both the negative and positive parts of the magnetic hysteresis, the coercivity varying linearly with the applied voltage. By assuming that the coercivity variations correspond to variations of the anisotropy energy[33], the efficiency of the electric field effect can be written $\eta = \frac{dE_k/S}{dE_{Gate}} = \frac{E_k/S}{H_C}\frac{dH_C}{dE_{Gate}}$, where $E_k/S$ is the anisotropy energy per surface, and $H_C$ the coercive field. In our system, $\eta$ reaches a value of -15 pJ/(V.m). This very large efficiency, which could be further enhanced by using thinner Mn$_4$N films, requires further in-depth studies to be understood. In particular, it is possible that a strain-based mechanism adds up to the charge transfer mechanism, due to the STO substrate behaviour in the low-temperature tetragonal phase[34,35,36]. Whatever the mechanism, this value is up to now only approached in systems with much thinner magnetic layers[37].

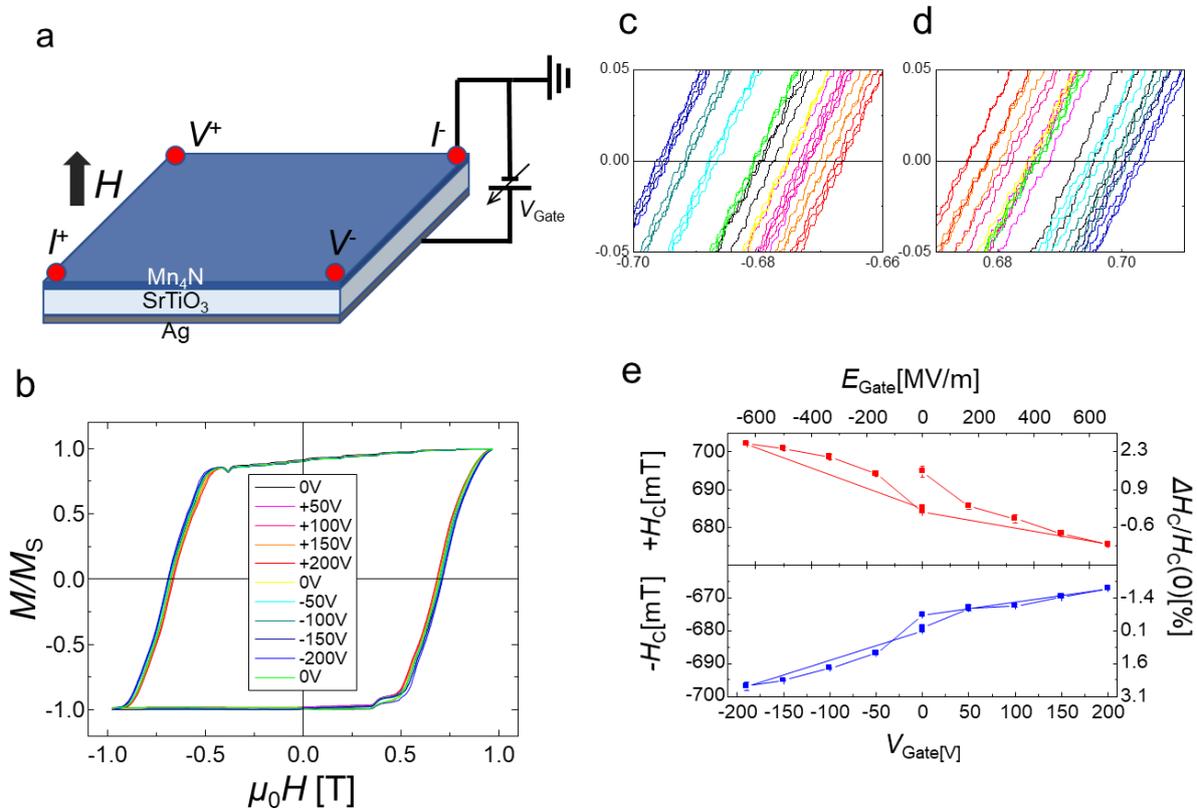

**Figure 4 | Voltage control of the coercive field of Mn$_4$N layers** (a) Schematics of the experiments. A gate voltage is applied to the back of the STO substrate, while the top electrical contacts, arranged in a Van der Pauw configuration, are used to measure the EHE of the Mn$_4$N layer. (b) Hysteresis loops recorded for various applied back gate voltages. (c) and (d) Zoom of the figure b close to the coercive field, for negative (c) and

positive (d) applied fields, using the same color code for the gate voltage. (d) Evolution of the coercivity with the applied gate voltage, for both the positive (top) and negative (bottom) half-loop.

**Conclusion**

In conclusion, $Mn_4N$/STO films present several properties that make them compelling candidates for spintronics applications. They are characterized by a perpendicular magnetisation, and notwithstanding the absence of heavy elements with high spin-orbit coupling, they exhibit a large extraordinary Hall effect. We showed in this system efficient current-induced domain wall motion, simply due to the adiabatic spin-transfer torque. The observed domain wall velocities largely surpass those obtained using spin transfer torques up to now, and is at least comparable to the best results reported for non-centrosymmetric stacks with interfacial DMI. We show that the low critical current is a consequence of the PMA and of the low magnetisation of the ferromagnetic material, which leads to a low barrier for precession of the DW magnetisation. The high DW mobility observed in the linear regime is a consequence of the small magnetisation and of the high spin polarisation. Finally, we demonstrate the possibility to control the magnetic anisotropy of $Mn_4N$ films using electric fields. As $Mn_4N$ is made of cheap and abundant elements, and does not include critical materials such as precious metals and rare-earths, it appears as a worthy candidate for sustainable spintronics applications. Future developments of this work may include the substitution of Mn by other 3d magnetic elements, in order to tailor the magnetic parameters, and for instance to reach the compensation point.

**Methods**

Sample growth

The $Mn_4N$ thin films were grown on STO substrates by plasma-assisted molecular beam epitaxy (MBE), as in refs.[6,38]. During the deposition, the substrate temperature was kept at 450 °C, to allow the diffusion of Mn and N atoms coming from the Mn solid source of a Knudsen-cell and from the radio-frequency (RF) $N_2$ plasma. We optimised the growth condition by using a Mn deposition rate of 1.5 nm/min, a $N_2$ gas flow of 1.0 $cm^3$/min, a pressure in the chamber of $4.5 \times 10^{-3}$ Pa, and an input power of 105 W. The samples were capped with $SiO_2$ for the CIDWM experiments, and with Ta for the voltage-control of the magnetic anisotropy experiments, using a sputtering gun installed in the MBE chamber.

XRD/RHEED

The $Mn_4N$ films were characterized *in-situ* by RHEED, with the 20 keV electron beam azimuth along the STO [100] direction. The diffraction pattern of anti-perovskite nitride consists of fundamental and superlattice diffraction peaks, similarly to what has been simulated for $Fe_4N$ thin film[39]. Our sample showed streaky RHEED patterns with obvious superlattice diffractions lines, which are sharper and surrounding the central line, indicating the presence of well ordered N atoms at the body center of the fcc Mn lattice.

The crystalline quality was also evaluated by $\omega$-$2\vartheta$ (out-of-plane), $\varphi$-$2\vartheta_\chi$ (in-plane) XRD and $\omega$-rocking curve measurement, using Cu K$\alpha$ radiation in a Rigaku Smart Lab®. For the in-plane diffraction, the incidence angle $\omega$ was fixed as 0.4°, and the scattering vector $Q$ was also set along STO [100] and [110]. The $\omega$-rocking curve at the Mn$_4$N[004] peak exhibits a remarkably small full width at half maximum of 0.14°, highlighting the very high crystalline quality of the epilayer[11]. The thickness of the layer was also measured by X-ray reflectometry.

AFM

The surface state was observed using a Bruker Dimension in the oscillating mode. The root mean square roughness (0.7 nm) has been evaluated on a 4×4 µm$^2$ area.

TEM

Scanning transmission electron microscopy (STEM) measurements have been carried out using a Cs-corrected FEI Themis at 200 keV. High angle annular dark field (HAADF)-STEM images were acquired using a convergence semiangle of 20 mrad and collecting scattering >65 mrad. The STEM specimen was prepared by the FIB lift-out technique using a FEI dual-beam Strata 400S at 30 kV.

VSM/SQUID

The magnetisation and magnetic anisotropy were measured by Quantum design MPMS® 3 with out-of-plane (1.5 T) and in-plane (7 T) magnetic fields. The anisotropic energy $K_u$ was calculated from the area enclosed in-between the out-of-plane and in-plane magnetisation curves.

EHE and voltage control of the anisotropy

Extraordinary Hall effect loops and resistivity measurements were performed using the Van der Pauw method for blanket layers at 10K, in a cryostat placed in-between the poles of a 1.2 T electromagnets. The 4 corners of the sample were wire-bonded using Al wires. A back gate voltage up to ±200 V was applied by sticking the sample to the sample holder with silver paint.

STT measurements

The Mn$_4$N films were processed into 1 and 2-µm-wide strips by electron beam lithography and Ar ion milling technique, using lifted-off Al patterns as etching masks. The contact electrodes consist in a Ti/Au/Ti stack. The domain wall velocities were measured using Magneto-Optical Kerr microscopy. After saturation of the magnetisation, high field pulses were applied in the reversed direction to nucleate a domain in the injection pad and to inject domain walls into the nanostrips. The domain walls were then displaced using current pulses generated by a voltage generator, able to provide voltage pulses up to 80V in amplitude and down to 500 ps in duration. The domain wall velocities were calculated by dividing the domain wall displacement by the normalized pulse length, measured by an oscilloscope connected in series with the sample. Further verification of the DW speeds was done for some current density values by measuring the DW displacement $\Delta$x for different current pulse durations $\Delta$t, and extracting the DW velocity from the slope of the curve $\Delta$x *vs*. $\Delta$t.

**Acknowledgments**

We acknowledge N. Mollard for the TEM specimen preparation. The devices were prepared in PTA platform from Grenoble, with partial support from the French RENATECH network. This work received support from the French ANR programme through projects OISO (ANR-17-CE24-0026-03), TOPRISE (ANR-16-CE24-0017) and the Laboratoire d'Excellence LANEF (ANR-10-LEBX-51-01). TG acknowledge JSAP support.

**Author contributions**

The experiments have been performed by T. G. and M. K., and the simulations by J. P. G. H. O. performed the TEM study. The work has been supervised by L.V., T.S., S.P., J.P.A. and J.V. All co-authors discussed the results, and wrote the manuscript.

# Mn$_4$N ferrimagnetic thin films for sustainable spintronics

T. Gushi[1,2], M. Jovičević Klug[3,÷], J. Peña Garcia[3], H. Okuno[4], J. Vogel[3], J.P. Attané[2], T. Suemasu[1], S. Pizzini*[,3], L. Vila[†,2]

1. Institute of Applied Physics, Graduate School of Pure and Applied Sciences, University of Tsukuba, Tsukuba, Ibaraki 305-8573, Japan

2. Univ. Grenoble Alpes, CEA, CNRS, Grenoble INP, INAC, SPINTEC, F-38000 Grenoble, France

3. Univ. Grenoble Alpes, CNRS, Institut Néel, F-38042 Grenoble, France

4. Univ. Grenoble Alpes, CEA, INAC, MEM, F-38000 Grenoble, France

*stefania.pizzini@neel.cnrs.fr

† Laurent.vila@cea.fr

÷Permanent address : Kiel University, Institute for Materials Science, Kaiserstraße 2, 24143 Kiel, Germany

## Absence of Spin Orbit Torques

Domain Walls (DWs) having a chiral Néel structure stabilized by interfacial Dzyaloshinskii-Moriya interaction (DMI) can be driven by Spin-Orbit Torques (SOTs), arising from the Spin Hall Effect or from Rashba effects [1]. In order to exclude the contribution of SOTs to the DW dynamics in our system, we have carried out measurements aiming at establishing the nature of the DW structure. The domain wall dynamics was driven by an out-of-plane magnetic field, in the presence of a continuous longitudinal in plane field $B_X$. The domain wall velocity in the direction of the in-plane field depends on the nature of the internal DW structure. On one hand, in the case of Bloch walls, the DW speed increases as a function of $B_X$, and is the same for fields of opposite signs (*i.e.*, the speed vs. $B_X$ curve is symmetric). On the other hand, the presence of DMI induces an asymmetry of the speed vs. $B_X$ curve, with a minimum speed when the $B_X$ field compensates the DMI field [2,3].

Examples of DW displacements in the absence and in the presence of an in-plane magnetic field of 400 mT are shown in Figure S1 (a,b). The white contrast in the Kerr images represents the displacement of the DW during the field pulse. This displacement is isotropic in the presence of the in-plane field, which indicates the absence of DMI. The domain wall speed vs. Bx curve is shown in Figure S1 (c).

## Field and current driven domain wall dynamics

Magnetisation dynamics in ferromagnetic materials is governed by the Landau-Lifshitz-Gilbert equation, with additional terms added to account for other interactions such as the STT. Following the description proposed by Thiaville et al. [4], the magnetisation dynamics is given by:

$$\frac{\partial \boldsymbol{m}}{\partial t} = \gamma_0 \boldsymbol{H}_{eff} \times \boldsymbol{m} + \alpha \boldsymbol{m} \times \frac{\partial \boldsymbol{m}}{\partial t} - (\boldsymbol{u} \cdot \boldsymbol{\nabla})\boldsymbol{m} + \beta \boldsymbol{m} \times [(\boldsymbol{u} \cdot \boldsymbol{\nabla})\boldsymbol{m}], \qquad (1)$$

where $\boldsymbol{m}$ is the unit vector along the local magnetisation, $\gamma_0 = \gamma \mu_0$ where $\gamma$ is the gyromagnetic ratio and $\mu_0$ the vacuum permeability, $\boldsymbol{H}_{eff}$ is the micromagnetic effective field, and $\alpha$ is the Gilbert damping constant. The spin-drift velocity $\boldsymbol{u}$ is parallel to the electron flow direction, with $\boldsymbol{u} = \frac{g\mu_B}{2eM_s}P$ where $g \approx 2$ is the free electron's Landé factor, $\mu_B$ the Bohr magneton, $e$ the electron charge, $P$ the current polarisation factor and $M_s$ the spontaneous magnetisation. Finally, $\beta$ is the non-adiabatic term, representing a second-order term of the STT.

Let us review now the DW dynamics features under an applied magnetic field and spin-polarized current using the 1D model. To describe the DW motion in the 1D model, two collective coordinates are chosen: the DW centre, $q$, and the tilt angle of the DW magnetisation out of the DW plane, $\varphi$. The DW profile is described by the following ansatz:

$$\theta(x,t) = 2 \arctan \exp\left(\frac{x - q(t)}{\Delta}\right), \qquad (2a)$$
$$\phi(x,t) = \varphi(t), \qquad (2b)$$

Where $\boldsymbol{m} = (\sin\theta \cos\phi, \sin\theta \sin\phi, \cos\theta)$. Using a Lagrangian approach [5,6] we can find the equations of motion under current (applied along +x) and magnetic field applied along the easy-axis z:

$$\frac{1}{\Delta}\dot{q} - \alpha\dot{\varphi} = \frac{u}{\Delta} + \frac{\mu_0 \gamma H_D}{2}\sin 2\phi, \qquad (3a)$$
$$\dot{\varphi} + \frac{\alpha}{\Delta}\dot{q} = \beta\frac{u}{\Delta} + \mu_0 \gamma H_{app}, \qquad (3b)$$

Here, $\Delta(\varphi) = \sqrt{\frac{A}{K_u + K_D(\varphi)}}$ is the DW width, where $A$ is the exchange stiffness, $K_u$ the uniaxial anisotropy, $K_D = \frac{\mu_0 M_s^2}{2}(N_x \sin^2\varphi + N_y \cos^2\varphi - N_z)$ ($N_i$ being the demagnetizing coefficients), and $H_D = \frac{2K_D}{\mu_0 M_s}$ is the DW demagnetizing field [4,7].

When a DW is driven by a magnetic field, the DW moves in the steady regime with a constant value of $\varphi$ up to a field called the Walker field ($H_W$). Above $H_W$, the DW moves by transforming continuously, i.e., $\dot{\varphi} \neq 0$ (from transverse to vortex DW for in-plane systems and from Bloch to Néel DW in out-of-plane systems). This continuous transformation results in a drop of the velocity until it reaches a second linear v vs. H regime, with reduced mobility.

When the DW is driven by a spin polarized current, in the adiabatic limit ($\beta = 0$) the DW moves continuously only when it can start precessing and align with the incoming spin polarisation. This occurs above a threshold critical current density $J_c$:

$$J_c = \frac{2e}{\hbar P}\Delta(\varphi)K_D, \qquad (4)$$

Through the dependence on $K_D$, $J_c$ depends on the value of the spontaneous magnetisation and on the geometry of the film strip [8]. In systems with in-plane magnetisation, the transverse anisotropy constant is given by $K_D = |K_z - K_y| \approx |K_z| = \frac{1}{2}\mu_0 M_s^2 - K_u$. On the other hand, in thin films with perpendicular magnetisation, the transverse anisotropy constant, which is related to the energy difference between a Bloch and a Néel DW, is given by $K_D = |K_x - K_y|$. For a thin strip of thickness $t$ and width $w$, the DW can be modeled as an ellipse and the demagnetizing factors can be approximated as $K_x \approx \frac{1}{2}\mu_0 M_s^2 \left(\frac{t}{t+\pi\Delta}\right)$, $K_y \approx \frac{1}{2}\mu_0 M_s^2 \left(\frac{t}{t+w}\right)$ [7]. For a 1 μm wide and 10 nm thick Mn$_4$N strip like in our experiments, we obtain $K_D \approx 1.1 \times 10^3$ J/m$^3$. This value is much lower than that obtained for a permalloy strip with the same geometry ($K_D \approx 0.5 \times 10^6$ J/m$^3$). From Equation 4, the critical current density for Mn$_4$N is of the order of $1.9 \times 10^{10}$ A/m$^2$ while the one for permalloy is at least a factor 100 larger.

In the adiabatic limit, for current densities above $J_c$, the DW starts precessing and moves at a velocity given by:

$$\boldsymbol{v}_{precession} = \frac{1}{1+\alpha^2}\sqrt{u^2 - u_c^2} \quad (5)$$

where $\boldsymbol{u_c}$ is the spin-drift velocity at the critical current density. When the critical current density is very small (as it is the case for systems with PMA and small $M_s$), the latter expression can be approximated:

$$\boldsymbol{v}_{precession} \approx \frac{1}{1+\alpha^2}|\boldsymbol{u}| = \frac{1}{1+\alpha^2}\frac{g\mu_B}{2eM_s}P\boldsymbol{J} \quad (6)$$

This expression then shows that the DW mobility in the precessional regime is proportional to the ratio between the spin polarisation and the spontaneous magnetization. Large mobilities are then expected for large polarisation and small $M_s$.

In order to confirm the different features predicted by the 1D model, we have performed micromagnetic simulations with the finite-difference software MuMax3 [9] in the pure adiabatic limit.

In order to limit the calculation time, the strip width was fixed to 120 nm and its thickness to 1 nm. We have compared the case of an in-plane magnetized system ($M_s = 1.4 \times 10^6 \frac{A}{m}$, $K_u = 9.75 \times 10^5$ J/m^3 and Q=0.79) with an out-of plane system ($M_s = 1.4 \times 10^6 \frac{A}{m}$, $K_u = 1.95 \times 10^6$ J/m^3 and Q= 1.51), where $Q = \frac{2K_u}{\mu_0 M_s^2}$ is the so-called quality factor. The obtained current-driven DW velocities are shown in Figure S2. For Q=1.51 an overall agreement is obtained with the 1D model: the critical current density is of the order of $10\, u \approx 2 \times 10^{10} A/m^2$ so that the DWs start moving linearly starting from very low current densities. Slight differences with respect to the 1D model appear for very low values of $u$ (cf. inset of Fig. S2): the DW moves discontinuously because of extrinsic effects, but it can nevertheless be displaced over some distance by the application of a current larger than $J_c$. Moreover, for high values of $J$, some deviations from the linear regime are observed. These deviations are associated to the DW asymmetry [10,11] and to the constraints on the magnetisation dynamics introduced by the use of the ansatz in the 1D model. The 1D model considers only 2 collective coordinates $(q, \varphi)$, whereas in a 2D extension of this model, the asymmetry is taken into account by considering an extra collective variable, $\chi$ [11].

When $Q < 1$, for in-plane magnetized systems, the high dipolar cost for bringing the magnetisation out-of-the-plane and into precession traduces into a high critical current (about $300\ u \approx 8 \times 10^{12} A/m^2$) as expected from Equation 4. The linear regime of motion is then obtained for much larger current densities.

# Influence of the non-adiabatic torque and of the damping parameter on the current-induced DW motion

Up to now we have neglected the effect of the non-adiabatic torque in the dynamics of DWs in PMA systems. While this term had to be considered to explain the DW motion observed experimentally in in-plane magnetized systems like permalloy strips below the intrinsic $J_c$, we have shown that the large DW velocities observed in our $Mn_4N$ strips for low current densities can be explained by the adiabatic term alone. It has been shown analytically that the non-adiabatic torque results in a steady regime motion with velocity $v = \frac{\beta}{\alpha} u$ below the critical current $J_c$. On the other hand, a term $\frac{\alpha\beta}{1+\alpha^2} u$ is added to the velocity in the precession regime [3]. However, since $\alpha$ is expected to be much smaller than 1 in $Mn_4N$ and $\beta$ is expected to be of the same order of amplitude as $\alpha$, this term may be neglected

In order justify our choice to neglect the non-adiabatic torque, we performed micromagnetic simulations considering two extreme cases: $\beta = 5 \cdot \alpha$ and $= 0.5 \cdot \alpha$, with $\alpha = 0.03$. The results are shown in Figure S3.

When the non-adiabatic torque is taken into account, the DW moves in the steady regime until it reaches the Walker spin-drift velocity:

$$u_W = u_c \cdot \frac{\alpha}{|\beta - \alpha|} \qquad (6)$$

as shown in Figure S3(a). Note that when the non-adiabatic torque is neglected, the Walker spin-drift velocity coincides with the critical spin-drift velocity. Beyond $J_c$, as described previously, the DW starts moving linearly with J, in the precessional regime, and the DW mobility is practically independent from the value of $\beta$ (Figure S3 (b)).

These results led us to conclude that the non-adiabatic torque is not necessary to explain the velocities obtained experimentally. Nevertheless, our experiments do not allow to exclude *a priori* the presence of the non-adiabatic term, because the regime where it expresses itself, below $J_c$, is hidden by the thermally activated regime.

The Gilbert damping plays a critical role in the DW motion. In our simulations we have considered a quite large damping $\alpha = 0.03$, but this value has not been verified experimentally. In order to evaluate the impact of the value of alpha to the DW mobility, we have carried out micromagnetic simulations for the case $\alpha = 0.03$ and $\alpha = 0.3$, and compared the results with the 1D model. The latter is the typical Gilbert damping value obtained for nm-thick Co layers deposited on high spin-orbit Pt layers. The results are shown in Figure S4. The difference in velocity for the maximum current density considered here is only about 6%. The larger damping also results into a larger $u_c$ but the difference is negligible (see inset of Figure S4). Therefore, we can conclude that considering $\alpha = 0.03$ does not modify strongly our results.

## Damping and non-adiabatic torque: the particular case for which α=β

So far, we have divided the DW motion behaviour in two regimes: the steady regime below the Walker breakdown, and the precessional regime for currents above the Walker breakdown. Simulating the DW motion in the precessional regime, and modelling our system as a low-damping material (which implies that $v_{precession} \approx u$) has allowed us to reproduce the experimental DW velocities.

However, when the Gilbert damping and the non-adiabatic torque compensate each other, *i.e.*, in the hypothesis where $\alpha = \beta$, the 1D model predicts that the Walker spin-drift velocity becomes infinite (Eq.6 ). In this particular case, the DW moves in the steady regime at a velocity equal to the spin-drift velocity $v_{steady} = u$, whatever the applied current density. In order to confirm this result, we have simulated a PMA strip with $Q = 1.51$, and $\alpha = \beta = 0.03$. The results are shown in Figure S5 (a) where they are compared with the case $\beta = 0$. We can note that the velocities do not vary significantly between the two cases. On the other hand, the DW moves as expected in the precessional regime when $\beta = 0$, while it moves in the steady flow regime when $\alpha = \beta$. This is confirmed by the temporal evolution of the averaged x- and y- components of the DW magnetisation that are plotted in Figure S5 (b-c).

Note that an experimental evidence of a case where $\alpha = \beta$ has been reported [12,13]. Here, owing to the small difference between the DW velocity when the non-adiabatic and the damping torque are balanced or unbalanced, we cannot distinguish experimentally between the two cases.

**Simulations details**

The micromagnetic simulations have been performed with the finite-difference software MuMax3 [9]. Zero-temperature simulations were performed in a defect-free strip of 6000×120×10 nm³ with a cell size of 2.5×2.5×10 nm³. For the study of the influence of $Q = \frac{2K_u}{\mu_0 M_S^2}$ on the DW motion, we set $M_S$=1.4×10⁶ A/m and we tuned $K_U$. For the simulations of the experiments we set $M_S$=7.1×10⁴ A/m and $K_U$=0.16×10⁶ J/m³. The rest of magnetic parameters were: A=10 pJ/m, P=0.7, α=0.03, β=0.

In a first step, the DW configuration in equilibrium was found in the absence of a spin-polarized current. In a second step, a current was applied along the positive x-axis, inducing an adiabatic torque resulting into the DW motion. We set up a post-step function that makes the simulation box "follow" the DW. The DW velocity was calculated by fitting the DW position as a function of time, 25 ns after the current was switched on, to ensure that all transient effects are damped out.

Finally, the influence of the cell size, of the damping parameter, of the DW width, of the non-adiabatic torque, and of the geometry were also tested. It was concluded that in the precession regime, none of these parameters strongly affects the DW motion.

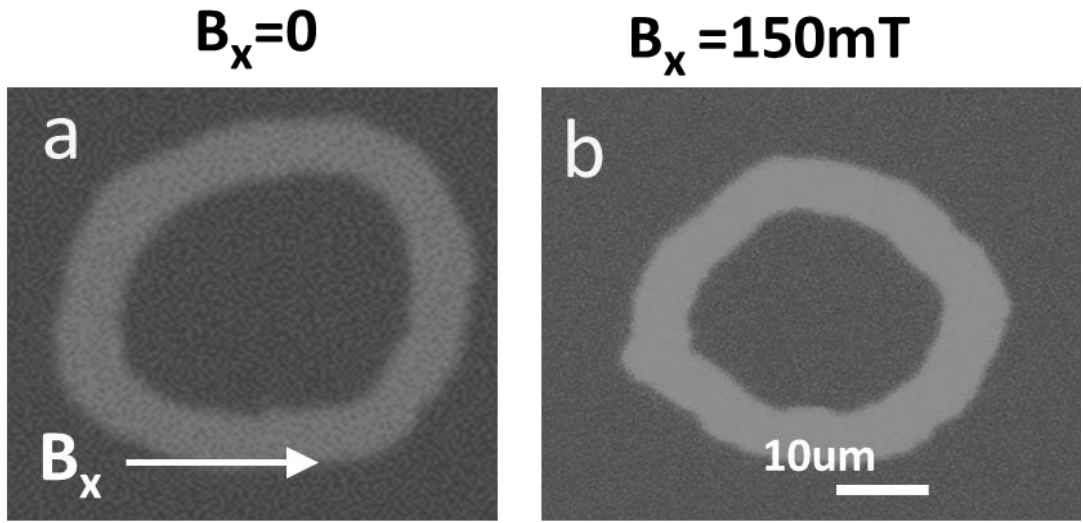

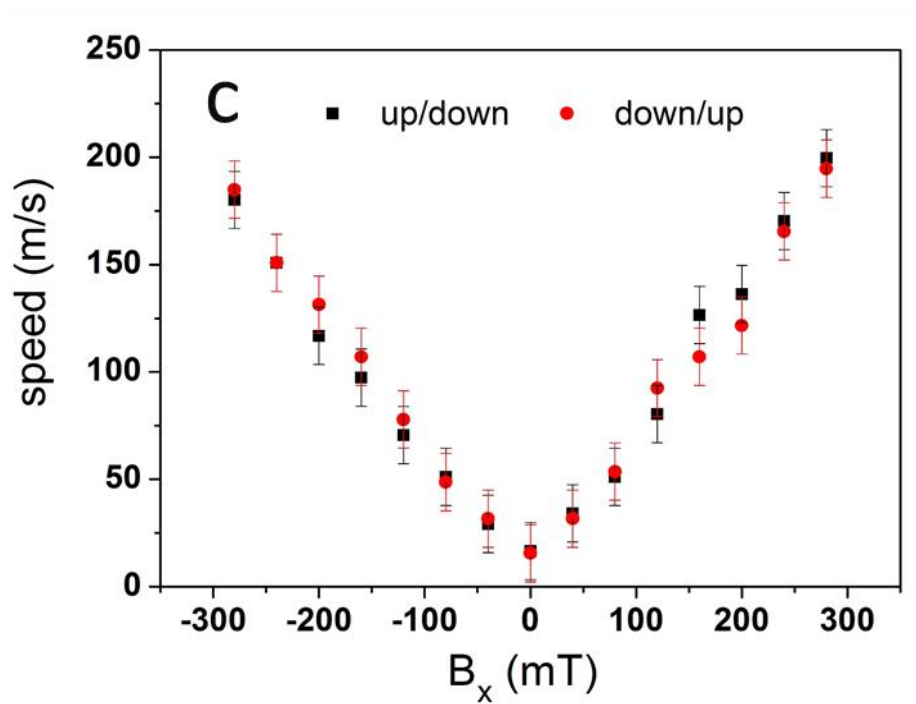

**Figure S1**. (a,b): Differential Kerr images showing the displacement of domain walls driven by an out-of-plane magnetic field pulse, of around 400mT and 30ns (white contrast) in the absence (a) and in the presence (b) of an in-plane continuous magnetic field $B_x$. (c): domain wall speed driven by $B_z$= 320 mT perpendicular field pulses, as a function of the in-plane field amplitude $B_x$, for an up/down and a down/up domain wall. The symmetric curve confirms the presence of achiral domain walls.

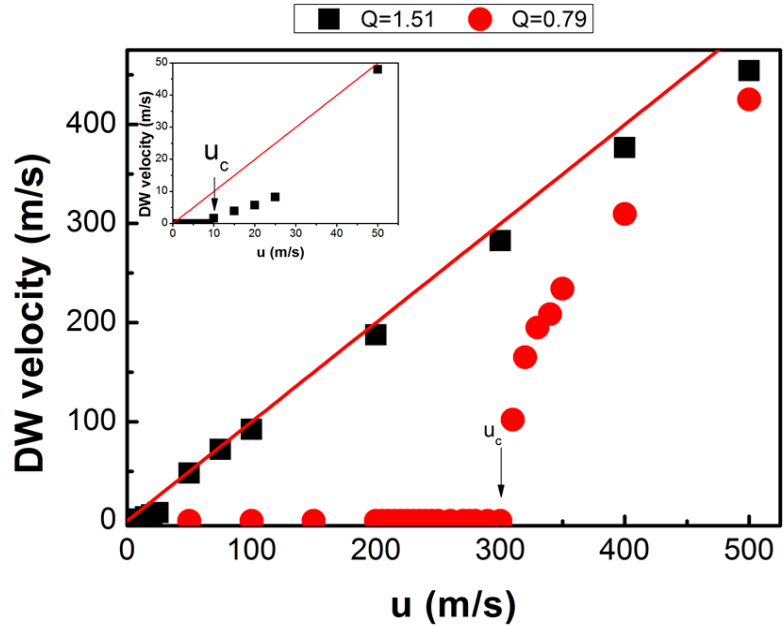

**Figure S2.** Micromagnetic simulations showing the DW velocity as a function of the spin-polarized current density, for two different values of $Q$: $Q = 1.51$ (black squares) and $Q = 0.79$ (red dots). The red solid lines represents the 1D model velocity (Eq. 6). The inset shows the details for small values of the spin-drift velocity.

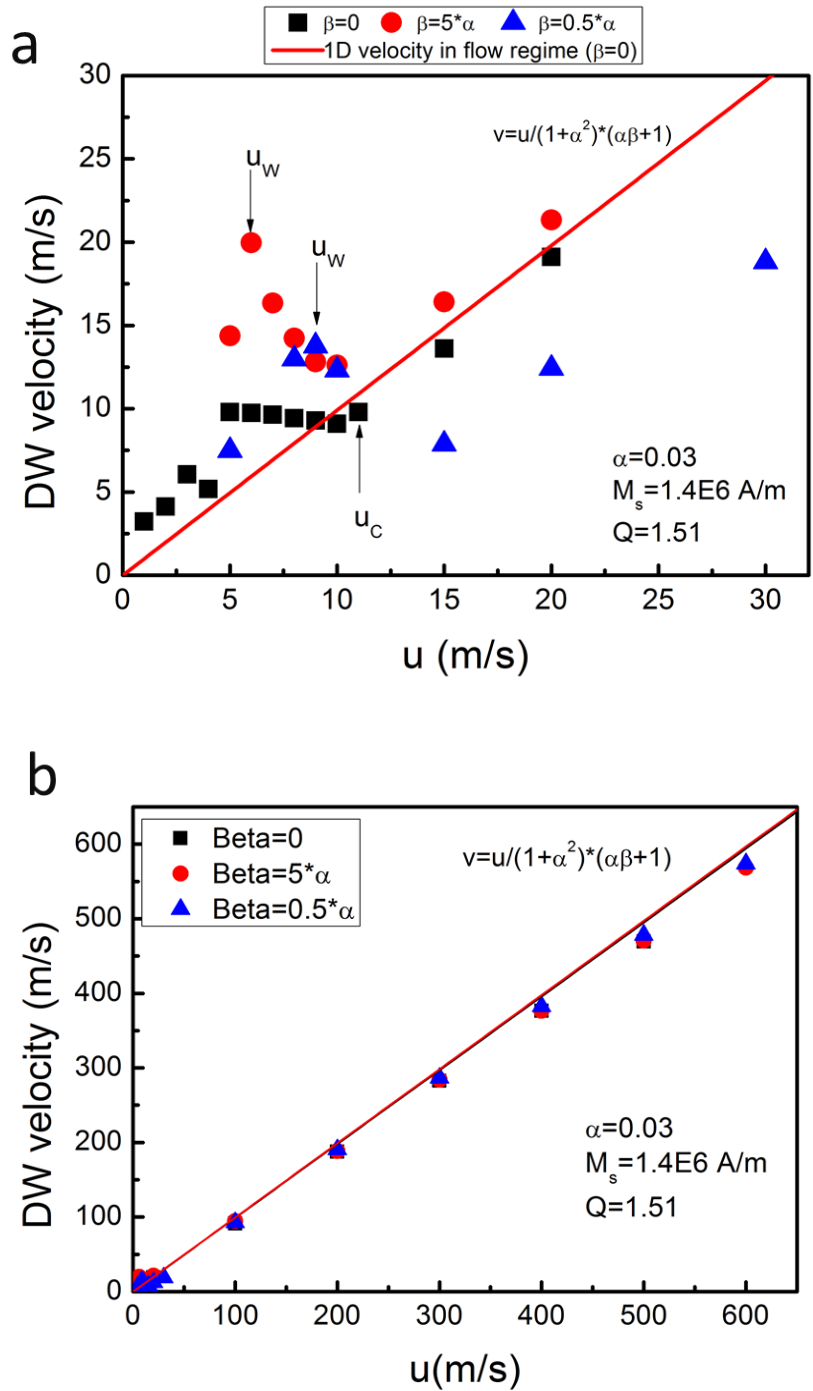

**Figure S3.** Results of micromagnetic simulations showing the DW velocity as a function of the spin-polarized current for different values of $\beta$: $\beta = 5\alpha$ (red circles), $\beta = 0.5\alpha$ (blue triangles) and $\beta = 0$ (black squares), for (a) low current and (b) large current densities. The red solid line represents the 1D model velocity (Eq. 5).

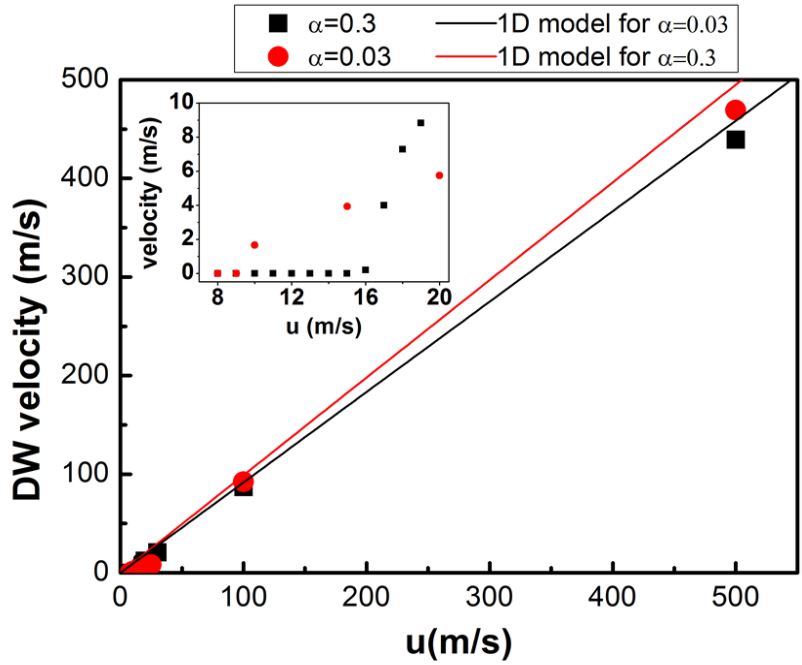

**Figure S4.** Results of micromagnetic simulations showing the DW velocity as a function of the spin-drift-velocity for two different values of $\alpha$: $\alpha = 0.03$ (red circles), $\beta = 0.3$ (black squares). The solid lines represent the 1D model velocity (Eq. 5).

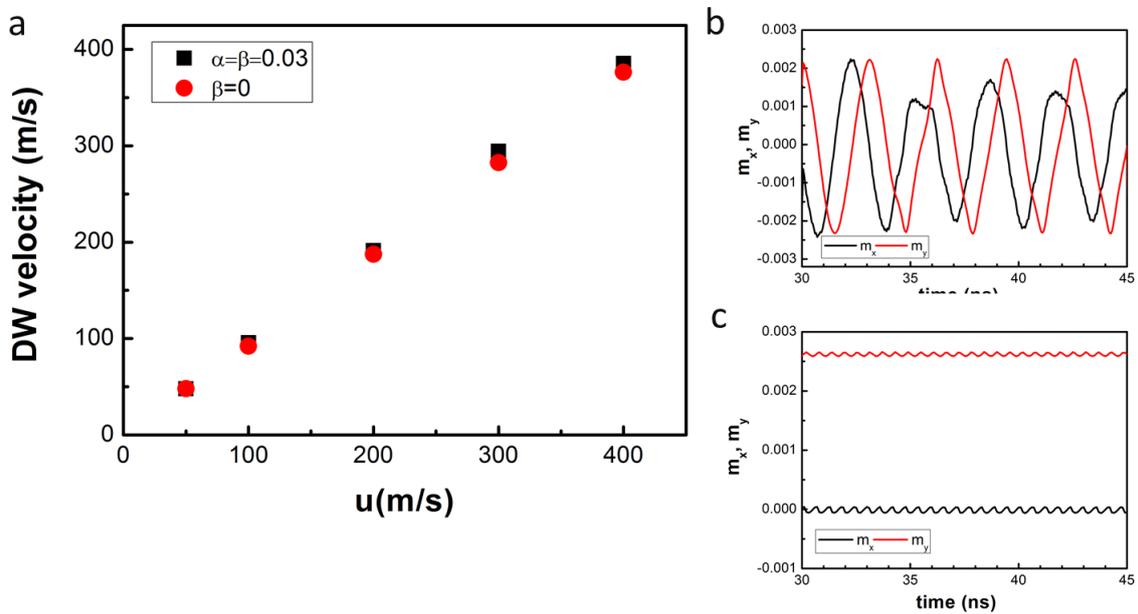

**Figure S5.** Results of micromagnetic simulations showing (a): the DW velocity as a function of the spin-drift velocity for $\beta = 0$ and $\beta = \alpha = 0.03$. (b,c): the temporal evolution of the averaged x- and y- DW magnetic components: (b) and $\beta = 0$ and (c) $\beta = \alpha = 0.03$ (c).